\documentclass[aps,pra,floatfix,twocolumn,showpacs]{revtex4}

\begin{document}

\title{Calculations of the spectra of superheavy elements E119 and E120$^+$}

\author{T.H. Dinh, V.A. Dzuba, V.V. Flambaum, and J.S.M. Ginges}

\affiliation{School of Physics, University of New South Wales,
	Sydney NSW 2052, Australia}
\date{\today}

\begin{abstract}

High-precision calculations of the energy levels of the superheavy elements E119 and E120$^+$ are presented. 
Dominating correlation corrections beyond relativistic Hartree-Fock are included to all orders in the 
Coulomb interaction using the Feynman diagram technique and the correlation potential method.
The Breit interaction and quantum electrodynamics radiative corrections are considered. Also, the 
volume isotope shift is determined.
A similar treatment for Cs, Fr, Ba$^+$ and Ra$^+$ is used to gauge the accuracy of the calculations and to 
refine the {\it ab initio} results.

\end{abstract}


\maketitle

\section{Introduction}

There has been great progress in recent years in the synthesis 
of superheavy elements (nuclear charge $Z>104$). 
Elements up to $Z=118$, excluding $Z=117$, have been produced (see, e.g., Refs. \cite{gsi,jinr}), 
and very recently evidence for  
naturally-occurring E122 was reported \cite{E122}.

Studies of superheavy elements are largely motivated by the predicted ``island of stability'' 
which occurs due to the stabilizing nuclear shell effects.
Different nuclear models vary in their predictions of the superheavy shell structure (see, e.g., \cite{scmf}). 
Experimental investigation of superheavy elements enables one to distinguish between different models. 

Experimental efforts are underway to measure the spectra and chemical properties of superheavy elements 
\cite{chemistry}. 
A number of theoretical works, from the quantum chemistry and atomic physics communities, have been devoted 
to these studies (see references in \cite{pershina_review} and \cite{eliav122}). 

Leading relativistic effects grow as $(Z\alpha)^2$, where 
$\alpha = e^2 /\hbar c$ is the fine structure constant, and they become very
large in superheavy elements. 
It has been shown that these effects lead to a number of interesting features,  
such as level inversion in the spectra of some elements ($s$-levels are pulled in 
and screen the Coulomb potential seen by higher-orbital waves such as $d$-waves, thereby 
pushing them out) \cite{pershina_review}.

In the present work we perform relativistic calculations to determine the spectra 
of superheavy elements E119 and E120$^+$. The isotope $^{292}$E120 is predicted to be doubly
magic in relativistic mean-field nuclear calculations.

\section{Method of calculation}

We perform calculations for Cs, Fr, Ba$^+$, and Ra$^+$ to help gauge the accuracy of the 
calculations for E119 and E120$^+$ and as a means to reduce the {\it ab initio} errors for the 
spectra of these elements through extrapolation.

At the first stage of the calculations we use the relativistic Hartree-Fock (RHF) method. 
Calculations are performed in the self-consistent potential formed by the $N-1$ electrons in the 
core ($V^{N-1}$ potential). A complete set of single-electron orbitals is obtained in this way. The 
orbitals satisfy the equation
\begin{equation}
h_{0} \psi_{0}=\epsilon_{0} \psi_{0} \ ,
\end{equation}
where $h_{0}$ is the relativistic Hartree-Fock Hamlitonian
\begin{equation}
\label{hf}
h_{0}=c\mbox{\boldmath$\alpha$}\cdot{\bf p}+(\beta-1)m c^{2}-\frac{Ze^{2}}{r}+V^{N-1} \ .
\end{equation}
Here $V^{N-1}=V_{\rm dir}+V_{\rm exch}$ is the sum of the direct and 
exchange Hartree-Fock potentials, $N$ is the number of electrons, $N-1$ is the number 
of electrons in the closed core, and $Z$ is the nuclear charge.

\subsection{Correlations} 

The main challenge in calculations of the spectra of superheavy elements is accurate 
treatment of correlations. We take into account correlations using the correlation potential 
method \cite{corrpot}. Here, a correlation potential operator $\Sigma$ is constructed 
such that its average value for the valence electron coincides with the correlation correction 
to the energy, $\delta\epsilon_{a}=\langle a |\Sigma |a \rangle$ 
     
When the single-particle orbitals are found in the Hartree-Fock potential, the many-body perturbation theory 
expansion for $\Sigma$ starts in second order in the Coulomb interaction. There are direct and exchange contributions to the 
correlation potential. Second-order $\Sigma$
is calculated via direct summation over a discrete set of single-particle 
orbitals. Rather than working with finite sums and integrals over the real spectrum, we use 
finite sums over a pseudo-spectrum. We introduce a cavity of radius $r= 40$~a.u 
and 40 B-splines are used as a basis for the functions.

The {\it ab initio} calculations may be improved by including three dominating higher-order 
diagrams into the second-order correlation potential \cite{higher}. These are 
(i) screening of the Coulomb interaction, (ii) the hole-particle interaction in the polarization operator,
and (iii) chaining of the correlation potential $\Sigma$.

In particular, (i) and (ii) are included into the direct diagrams of $\Sigma$ using the Feynman diagram technique. 
For the exchange diagrams we use factors in the second-order $\Sigma$ to imitate the effects of 
screening. These factors are $f_{0}= 0.72$, $f_{1}= 0.62$, $f_{2}= 0.83$, $f_{3}= 0.89$, $f_{4}= 0.94$, $f_{5}= 1$; 
the subscript denotes the multipolarity of the Coulomb interaction. These factors have been estimated 
from accurate calculations of the higher-order corrections. 
The chaining of the correlation potential (iii) is included trivially by adding $\Sigma$ to the 
Hartree-Fock potential. The energies, with correlations included, are solutions of the equations for the valence 
electrons,
\begin{equation}
(h_{0}+\Sigma) \psi_{a}=\epsilon_{a} \psi_{a} \ .
\end{equation}

Further improvements to the wave functions and energies may be made semi-empirically through the use of fitting factors $f$  
(not to be confused with the Coulomb screening factors above) placed before the correlation potential, i.e.,
\begin{equation}
(h_{0}+f\Sigma) \psi_{a}^\prime=\epsilon_{a}^\prime \psi_{a}^\prime \ .
\end{equation}
Factors used for E119 (E120$^+$) are found by fitting to the experimental energies of the lighter electronic analogs 
Cs and Fr (Ba$^+$ and Ra$^+$). The use of fitting factors is considered a means of including effects, such as higher-order 
correlations, beyond what is included in the {\it ab initio} approach. 
 
\subsection{Breit interaction}

We go beyond the treatment of the electron-electron interaction in the Coulomb approximation, taking into account 
magnetic and retardation effects through inclusion of the Breit interaction. We use the following form for the Breit operator
\begin{equation}
h^{B}=-\frac{\mbox{\boldmath$\alpha$}_{1}\cdot \mbox{\boldmath$\alpha$}_{2}+
(\mbox{\boldmath$\alpha$}_{1}\cdot {\bf n})
(\mbox{\boldmath$\alpha$}_{2}\cdot {\bf n})}{2r} \ ,
\end{equation}  
where ${\bf r}={\bf n}r$, $r$ is the distance between electrons, and 
$\mbox{\boldmath$\alpha$}$ is the Dirac matrix.

In a similar way to the Coulomb interaction, we determine the self-consistent 
Hartree-Fock contribution arising from Breit. This is found by solving 
Eq. (\ref{hf}) in the potential
\begin{equation}
V^{N-1}=V^{C}+V^{B} \ ,
\end{equation}  
where $V^{C}$ is the Coulomb potential, $V^B$ is the Breit potential.

\subsection{Lamb shift}

Quantum electrodynamics radiative corrections to the energies (Lamb shifts) are accounted for by use of the 
radiative potential introduced in Ref. \cite{radpot}. This potential has the form
\begin{equation}
V_{\rm rad}(r)=V_U(r)+V_g(r)+ V_f(r) + V_l(r) \ ,
\end{equation}
where $V_U$ is the Uehling potential and $V_g$ is the potential arising from the magnetic 
formfactor. The potential corresponding to the electric formfactor is divided into low- and high-frequency parts, 
respectively: 
\begin{equation}
\label{low}
V _{l}(r)= -\frac{B(Z)}{e}Z^4 \alpha^5 m c^2 {\rm e}^{-Zr/a_B} 
\end{equation}
and 
\begin{eqnarray}
\label{high}
&&V _{f}(r)=-A(Z,r)\frac{\alpha }{\pi}V (r)
\int _{1}^{\infty}dt~ \frac{1}{\sqrt{t^2-1}}
\Big[\Big(1- \frac{1}{2t^2}\Big)\times \nonumber \\
&& \Big(\ln{(t^2-1)}+4\ln{(\frac{1}{Z\alpha}+0.5)}\Big)
-\frac{3}{2}+\frac{1}{t^2}\Big] 
{\rm e}^{-2trm} \ ;   
\end{eqnarray}
$V(r)$ is the nuclear potential, the coefficient  $A(Z,r)=(1.071-1.976x^2-2.128x^3+0.169x^4)
mr/(mr+0.07Z^2\alpha ^2)$, where $x=(Z-80)\alpha$, 
and $a_B$ is the Bohr radius.
Eqs. (\ref{low},\ref{high}) were determined semi-empirically by fitting to the Lamb shifts of high states 
of hydrogen-like ions for Z=10-110.

This potential is added to the 
Hartree-Fock potential,
\begin{equation}
V^{N-1}=V^{N-1}+V_{\rm rad} \ .
\end{equation} 
It is included in the self-consistent solution of the core Hartree-Fock states. Core relaxation, demonstrated 
to be important for the energies of valence $p$-states, is therefore taken into account.

\section{Results and Discussion}

We have calculated removal energies for the low-lying states $s$, $p_{1/2}$, and $p_{3/2}$.  
Results for Cs, Fr, and E119 are presented in Table \ref{tab:E119} and those for 
the ions Ba$^+$, Ra$^+$, and E120$^+$ are presented in Table \ref{tab:E120}. We list results in the RHF 
approximation and those with correlations included (with dominant diagrams summed to all orders). 
The {\it ab initio} results are listed under the column ``$\Sigma$''. In the column to the right, the percentage 
deviation from experiment is given in brackets. It is seen for Cs, Fr, Ba$^+$, and Ra$^+$ 
that there is excellent agreement with experiment, with disagreement on the order of $0.1\%$. 
The largest disagreements are for $7p_{1/2}$ for both Fr ($0.5\%$) and Ra$^+$ ($0.4\%$).

\begin{table}
\caption{Removal energies for states of Cs, Fr, and E119 (cm$^{-1}$). 
$f_{\rm Cs}$, $f_{\rm Fr}$ are factors placed before $\Sigma$, found by fitting to the energies of Cs and Fr, respectively. 
Numbers in brackets are percentage deviations of the preceding result compared to experiment. 
In the last column for E119 we present results of another high-precision calculation, Ref. \cite{eliav}.
Final values in the present work for E119 are in column ``$f_{\rm Fr}\Sigma$''.}
\label{tab:E119}
\begin{ruledtabular}
\begin{tabular}{l l r r r r r r r}
Atom & State & RHF & $\Sigma$ & & $f_{\rm Cs}\Sigma$ & & $f_{\rm Fr}\Sigma$ & Exp.\footnotemark[1]/ \\
 & & & & & & & & Other\footnotemark[2]\\
\hline
Cs & $6s$       & 27954 & 31467 & (0.2) & & & & 31407 \\
&$7s$       & 12112 & 12873 & (0.0) && &  & 12872 \\
&$8s$       &  6793 &  7090 & (0.0) && &  & 7090 \\
&$6p_{1/2}$ & 18791 & 20295 & (0.3) && &  & 20228 \\
&$7p_{1/2}$ &  9223 &  9662 & (0.2) && &  & 9641 \\
&$8p_{1/2}$ &  5513 &  5707 & (0.2) & & & & 5698 \\
&$6p_{3/2}$ & 18389 & 19727 & (0.3) & & & & 19674 \\
&$7p_{3/2}$ &  9079 &  9478 & (0.2) & & & & 9460 \\
&$8p_{3/2}$ &  5446 &  5623 & (0.1) & & & & 5615 \\

Fr &$7s$       & 28768 & 32931 & (0.2)  & 32860 & (0.0) & & 32849 \\
&$8s$       & 12282 & 13116 & (0.1) &  13115 & (0.0) &  & 13109 \\
&$9s$       &  6858 &  7177 & (0.0) & 7177 & (0.0) & & 7178 \\
&$7p_{1/2}$ & 18855 & 20708 & (0.5) & 20625 & (0.1) && 20612 \\
&$8p_{1/2}$ &  9240 &  9762 & (0.3) & 9737 & (0.0) &&  9736 \\
&$9p_{1/2}$ &  5521 &  5747 & (0.3) & 5738 & (0.1) && 5731\\
&$7p_{3/2}$ & 17655 & 18970 & (0.2) & 18919 & (0.0) && 18925 \\
&$8p_{3/2}$ &  8811 &  9206 & (0.2) & 9189 & (0.0) && 9191 \\
&$9p_{3/2}$ &  5319 &  5494 & (0.2) & 5487 & (0.1) && 5483\\

E119 & $8s$  & 33554 & 38954 && 38866 && 38852 & 38577 \\
& $9s$       & 13194 & 14087 && 14086 && 14079 & 14050 \\
& $10s$      &  7208 &  7534 && 7535 && 7536 & 7519 \\
& $8p_{1/2}$ & 20126 & 23445 && 23294 && 23272 & 22979 \\
& $9p_{1/2}$ &  9654 & 10453 && 10416 && 10415 & 10365 \\
&$10p_{1/2}$ &  5709 &  6040 &&  6027 && 6018 & 5997 \\
&$8p_{3/2}$  & 16674 & 18102 && 18046 && 18053 & 18007 \\
&$9p_{3/2}$  &  8449 &  8883 &&  8863 && 8866 & 8855 \\
&$10p_{3/2}$ &  5145 &  5340 &&  5332 && 5328 & 5320 \\

\end{tabular}
\footnotetext[1]{Cs data from Ref. \cite{moore} and Fr data from Ref. \cite{NIST}.}
\footnotetext[2]{Values for E119 are results of calculations, Ref. \cite{eliav}.}
\end{ruledtabular}
\end{table}

\begin{table}
\caption{Removal energies for levels of Ba$^{+}$, Ra$^{+}$, and E120$^{+}$ (cm$^{-1}$).
$f_{\rm Ba}$, $f_{\rm Ra}$ are factors placed before $\Sigma$, found by fitting to the energies of Ba$^{+}$ and Ra$^{+}$, respectively. 
Numbers in brackets are percentage deviations of the preceding result compared to experiment.  
Final values for E120$^+$ are in column ``$f_{\rm Ra}\Sigma$''.}
\label{tab:E120}
\begin{ruledtabular}
\begin{tabular}{l l r r r r r r r}
Atom & State & RHF & $\Sigma$ & & $f_{\rm Ba}\Sigma$ & & $f_{\rm Ra}\Sigma$ & Exp.\tablenotemark[1] \\
\hline
Ba$^+$ &$6s$& 75340 & 80834 & (0.2) &&& & 80686 \\
&$7s$       & 36852 & 38344 & (0.0) &&& & 38331 \\
&$8s$       & 22023 & 22662 & (0.0) &&& & 22661 \\
&$6p_{1/2}$ & 57266 & 60603 & (0.3) &&& & 60425 \\
&$7p_{1/2}$ & 30240 & 31346 & (0.2) &&& & 31296 \\
&$8p_{1/2}$ & 18848 & 19365 & (0.1) &&& & 19350 \\
&$6p_{3/2}$ & 55873 & 58876 & (0.2) &&& & 58734 \\
&$7p_{3/2}$ & 29699 & 30718 & (0.1) &&& & 30675 \\
&$8p_{3/2}$ & 18580 & 19060 & (0.1) &&& & 19050 \\

Ra$^+$ &$7s$& 75899 & 82034 & (0.2) & 81871 & (0.0) & & 81842 \\
&$8s$       & 36861 & 38454 & (0.0) & 38440 & (0.0) & & 38437 \\
&$9s$       & 22005 & 22677 & (0.0) & 22675 & (0.0) & & 22677 \\
&$7p_{1/2}$ & 56878 & 60743 & (0.4) & 60535 & (0.1) & &  60491 \\
&$8p_{1/2}$ & 30053 & 31297 & (0.2) & 31241 & (0.0) & & 31236 \\
&$9p_{1/2}$ & 18748 & 19322 &  & 19306 & & &       \\
&$7p_{3/2}$ & 52906 & 55771 & (0.2) & 55634 & (0.0) & & 55634 \\
&$8p_{3/2}$ & 28502 & 29493 & (0.1) & 29451 & (0.0) & & 29450 \\
&$9p_{3/2}$ & 17975 & 18445 & (0.1) & 18436 & (0.0) & & 18432 \\

E120$^+$ & $8s$ & 83168 & 90145 && 89964 && 89931 &\\
&$9s$        & 38468 & 40128 && 40113 && 40110 &\\
&$10s$       & 22673 & 23357 && 23355 && 23357 &\\
&$8p_{1/2}$  & 60027 & 65430 && 65141 && 65080  &\\
&$9p_{1/2}$  & 31121 & 32678 && 32609 && 32604 &\\
&$10p_{1/2}$ & 19253 & 19945 && 19926 && 19926\footnotemark[2] &\\
&$8p_{3/2}$  & 49295 & 52003 && 51873 && 51874 &\\
&$9p_{3/2}$  & 27028 & 27993 && 27952 && 27951 &\\
&$10p_{3/2}$ & 17223 & 17691 && 17681 && 17678 &\\
\end{tabular}
\footnotetext[1]{Ref. \cite{moore}.}
\footnotetext[2]{Final result corresponds to fitting from Ba$^+$, since there is no experimental data for Ra$^+$.}
\end{ruledtabular}
\end{table}

In the column ``$f_{\rm Cs}\Sigma$'' in Table \ref{tab:E119} we list the results for calculations for Fr and E119 
with the factor $f_{\rm Cs}$ found by fitting to the measured energies for Cs. It is clear by looking at the results 
for Fr that in all cases the results are significantly improved. The deviations from experiment are 0.1\% or better. 
It is the same situation for the results for Ra$^+$, as can be seen from Table \ref{tab:E120}.

From the trend in the corrections from Cs to Fr, we expect that using fitting factors significantly 
improves the accuracy of calculations for E119. Because use of the fitting factors $f_{\rm Cs}$ for Fr calculations 
leads to such good agreement with experiment, the fitting factors $f_{\rm Fr}$ differ only slightly from $f_{\rm Cs}$. This 
means that extrapolation of the spectra for E119 from Fr gives energies that are only slightly different from those found 
from extrapolation from Cs. Our final results for E119 are found using $f_{\rm Fr}$, presented in Table \ref{tab:E119}. 
In the final column of Table \ref{tab:E119} we list for E119 results of another high-precision calculation 
\cite{eliav} and postpone discussion of this work till Section \ref{sec:comparison}.

We see the same pattern for the ions, and our final results for E120$^+$ are listed under the column ``$f_{\rm Ra}\Sigma$''.

\subsection{Breit and radiative corrections}

Breit corrections were calculated in the self-consistent Breit-Hartree-Fock potential and the results 
are presented in Table \ref{tab:breit}. These numbers should be considered only as an indication of the 
order of magnitude of the corrections since the correlated Breit corrections may be large. For example, in 
Ref. \cite{derevianko_breit} it was found for Cs that account of correlations changes the sign for $6s$ 
(from $3.2\,$cm$^{-1}$ to $-2.6\,$cm$^{-1}$). In that work it was found that there is a small suppression 
due to correlations for $6p_{1/2}$ ($7.5\,$cm$^{-1}$ to $7.1\,$cm$^{-1}$) and for $6p_{3/2}$ it is significant 
($2.9\,$cm$^{-1}$ to $0.84\,$cm$^{-1}$).

\begin{table}
\caption{Corrections to removal energies from account of the Breit interaction. $n$ is the principal quantum 
number of the ground state. Units are $-$cm$^{-1}$.}
\label{tab:breit}
\begin{ruledtabular}
\begin{tabular}{l r r r r r r}
State          & Cs & Fr & E119 & Ba$^+$ & Ra$^+$ & E120$^+$ \\
\hline
$ns$           & 3 & 6 & 35 & 14 & 26 & 82 \\
$(n+1)s$       & 1 & 2 & 8 & 5 & 10 & 24 \\
$(n+2)s$       & 0 & 1 & 3 & 2 & 5 & 11 \\
$np_{1/2}$      & 7 & 14 & 34 & 29 & 52 & 112 \\
$(n+1)p_{1/2}$  & 3 & 5 & 11 & 11 & 20 & 41 \\
$(n+2)p_{1/2}$  & 1 & 2 & 5 & 6 & 10 & 20 \\
$np_{3/2}$      & 3 & 4 & 5 & 12 & 18 & 19 \\
$(n+1)p_{3/2}$  & 1 & 2 & 2 & 5 & 7 & 8 \\
$(n+2)p_{3/2}$  & 0 & 1 & 1 & 3 & 4 & 4 \\
\end{tabular}
\end{ruledtabular}
\end{table}

Our results for quantum electrodynamics (QED) radiative corrections are listed in Table \ref{tab:rad} 
alongside results of other calculations. We're not aware of other data for the ions. 
As with the Breit corrections, our results should only be considered estimates, to give an idea of the 
size of these corrections. They are calculated at the Hartree-Fock level, with correlation corrections 
excluded. (The effect of correlations would be to increase the density of the valence electrons at the 
nucleus, thereby leading to larger radiative corrections.)

Moreover, the radiative potential itself was found by fitting to states of hydrogen-like atoms for 
$10\le Z \le 110$. Due to a lack of data, direct fitting for $Z=119,\,120$ was not possible, and 
it is not clear how well our radiative potential would work in this region.

\begin{table}
\caption{Radiative corrections to removal energies. Units are $-$cm$^{-1}$.}
\label{tab:rad}
\begin{ruledtabular}
\begin{tabular}{l l c c c c c c}
Atom & State & This & Ref. & Ref. & Ref. & Ref. & Ref. \\
     &       & work & \cite{pyykko98} & \cite{labzowsky} & \cite{sapirstein} & 
\cite{pyykko03}\tablenotemark[1] & \cite{eliav} \\
\hline
Cs & $6s$ & 16 & 15.5 & 14.9$\to$26.6 & 12.7$\to$23.1 & 14.1  & 18.0 \\
   & $7s$ & 4  &      &                        & & & 4.2 \\
Fr & $7s$ & 36 & 38.3 & 37.1$\to$61.1 & 23.9$\to$52.6 & 40.6 & 28.8 \\
   & $8s$ & 9  &      &                        & & & 2.9 \\
E119 & $8s$ &  67& 141\tablenotemark[2] & 140$\to$152 & & 139  & 83.2 \\
   & $9s$   & 13 & &                           & & & 22.6 \\
   & $8p_{1/2}$ & 1 & & & & & 18.2 \\
   & $8p_{3/2}$ & 2 & & & & & 3.7 \\
Ba$^+$ & $6s$ & 37 & & & & & \\
   & $7s$ & 12 & & & & &\\
Ra$^+$ & $7s$ & 77 & & & & &\\
   & $8s$ & 24 & & & & &\\
E120$^+$ & $8s$ & 120 & & & & &\\
   & $9s$   & 32 & & &  && \\
& $8p_{1/2}$ & 5 & & & & &\\
   & $8p_{3/2}$ & 7 & & & & &\\
\end{tabular}
\tablenotetext[1]{Self-energies are given in Ref. \cite{pyykko03}; we have added vacuum polarization 
contributions from Ref. \cite{labzowsky} calculated at the Dirac-Fock level.}
\tablenotetext[2]{This number is quoted in their later work Ref. \cite{labzowsky} without explanation; 
in the original work Ref. \cite{pyykko98} the value is $211$ in the same units.}
\end{ruledtabular}
\end{table}

In Refs. \cite{pyykko98,eliav}, ratio methods were used to evaluate the self-energies.
In the former they were found from the ratio $E_{SE}\,\langle V_{VP}\rangle _{DF} /E_{VP}$, where 
$\langle V_{VP}\rangle_{DF}$ is the Uehling potential averaged over Dirac-Fock wave functions for the neutral system, 
and $E_{SE}$ and $E_{VP}$ are self-energy and vacuum polarization (Uehling) corrections to the energies in hydrogen-like 
systems. In the latter, the ratio is 
$E_{SE}\langle \nabla U_{\rm nuc}(r)\rangle_{DF}/\langle \nabla U_{\rm nuc}(r)\rangle_{\rm H}$, where $U_{\rm nuc}$ is the 
nuclear potential and $\langle\,\rangle_{H}$ denotes averaging over H-like states. 
In Refs. \cite{labzowsky,sapirstein} the Lamb shifts are found employing rigorous QED 
in the field of several different effective atomic potentials. The numbers in the tables 
give the ranges in the values for the potentials considered. Similarly to the current work, in Ref. \cite{pyykko03} 
an effective local potential, mimicking self-energy QED effects, is added to the Dirac-Fock potential. 

We see good agreement for the ``lighter'' atoms, though some disagreement for E119. 
We already mentioned why our results should be considered as order of magnitude estimates only for the superheavy 
elements.

We note that in our work, unlike in all other works mentioned, core relaxation is taken into account. 
This is accomplished by including the radiative potential into the self-consistent procedure for the core. 
While this effect is relatively small for Cs $s$ levels, it is significant for E119. For E119 8$s$, the Lamb 
shift changes from $85{\rm cm}^{-1}$ to $67{\rm cm}^{-1}$ without and with core relaxation, respectively. 
For $p$ levels the correction is more dramatic, although the size of the effect itself is much smaller. 
For E119 $8p_{1/2}$, we find the radiative correction to the binding energy without and with core relaxation to 
be $7{\rm cm}^{-1}$ and $1{\rm cm}^{-1}$. The effect of the core relaxation is to repel the inner electrons 
(the Lamb shift decreases the binding energy), leading to reduced shielding of the nuclear Coulomb field at 
small distances where the radiative corrections are determined. 

\subsection{Comparison with other calculations}
\label{sec:comparison}

We know of only one other work where high-precision calculations have been performed for spectra of the 
superheavy elements studied in this work. Eliav {\it et al.} \cite{eliav} have performed coupled cluster calculations 
for E119 spectra, including both Breit and radiative corrections. The results of their {\it ab initio} calculations 
are tabulated alongside our final (semi-empirical) values in Table \ref{tab:E119}. We see that generally there 
is agreement on the level $\sim 0.1\%$, with larger deviations for $8s$ (0.7\%) and $8p_{1/2}$ (1.3\%).

We investigated the large deviations for levels $8s$ and $8p_{1/2}$ by calculating the spectra of Cs, Fr, and E119 
with correlations calculated in the second order of perturbation theory ($\Sigma^{(2)}$,  
with no higher-order screening or hole-particle 
interactions taken into account). We used fitting factors to mimic higher-order effects, as was done with the full 
correlation potential $\Sigma$, and compared the results to those in Table \ref{tab:E119}. The 
result for $8s$ obtained in second order from the fit to Cs, $f_{\rm Cs}^{\prime}\Sigma^{(2)}_{\rm E119}$, differs from that 
in all orders, $f_{\rm Cs}\Sigma_{\rm E119}$, by 0.3\% (more bound) and for $8p_{1/2}$ the difference is
-0.3\%. With fitting to Fr spectra (using $f^{\prime}_{\rm Fr}$ and $f_{\rm Fr}$, respectively, before  
$\Sigma^{(2)}$ and $\Sigma$) the difference is 0.0\% for $8s$ and -0.3\% for $8p_{1/2}$. For other states, 
the agreement is -0.1\% or better. 

For E120$^+$, very good agreement for $s$-levels was obtained using the two approaches with fitting to 
Ba$^+$ and Ra$^+$ (0.1\% or better). 
For the $p$-levels, there are larger deviations, the largest being 0.5\% for $8p_{1/2}$. 

Differences in values for the spectra obtained in the two approaches gives an indication of the error 
from missed higher-order effects. An estimate of $\sim0.1\%$ error supports the detailed consideration below.

\subsection{Estimate of the accuracy}

Our final results for the superheavy elements do not include either Breit or radiative corrections and are 
listed in Tables \ref{tab:E119} and \ref{tab:E120}. The reason is that by using factors obtained by fitting 
to measured spectra, it appears that some of the Breit and radiative corrections are included. 

Let us consider states of Fr and Ra$^+$ with sizeable ($\sim 0.1\%$) Breit and radiative corrections. 
For Fr this is $7s$. At the {\it ab initio} level, the deviation from experiment is $0.25\%$ (column ``$\Sigma$''). 
With the fitting factor $f_{\rm Cs}$, the deviation is reduced to a tiny $0.03\%$. Moreover, the (estimated) 
contribution from Breit and radiative corrections (Tables \ref{tab:breit} and \ref{tab:rad}) is much larger than 
this deviation, being 0.13\% of the measured energy. For Ra$^+$, looking at energies for the states $7s$, $8s$, 
$7p_{1/2}$, $8p_{1/2}$, it is seen that we have the same story: the value obtained 
from fitting is everywhere better than the estimated Breit and radiative contributions. This strongly 
supports the argument that the use of empirical fitting factors takes into account not only the effects 
of higher-order correlation effects, but also the Breit and radiative corrections to some extent. 

The question then becomes: can we expect the same accuracy for E119 and E120$^+$ as has been demonstrated 
for Fr and Ra$^+$? Calculations for E119 and E120$^+$ were performed in a similar way as for Fr and Ra$^+$ 
and so we expect that extrapolation from the lighter to the heavier systems follows the same pattern we 
saw from Cs and Ba$^+$ to Fr and Ra$^+$. However, when we go to the heavier systems, there is some difference. 
For instance, the $Z$-dependence of the relativistic, Breit, and radiative corrections for the ``light'' systems 
is $\sim Z^2$, while for the superheavy elements this dependence is stronger \cite{pyykko98}. 
It means that extrapolation from Fr to E119, for instance, is probably not as good as extrapolation from Cs to Fr. 

For Cs, Fr, Ba$^+$, and Ra$^+$ it is seen that the largest uncertainty in the {\it ab initio} calculations 
comes from the unaccounted correlation corrections, these being larger than the estimated Breit and 
radiative corrections. This is the case for the higher states for E119 and E120$^+$, however for the ground state 
these corrections are about the same (0.3\% and 0.2\%, respectively).\footnote{A ballpark error for the 
{\it ab initio} correlation calculations is found by looking at the difference of the correlated values with and 
without fitting.} We expect that, as with Fr and Ra$^+$, use of the empirical fitting factors improves the accuracy 
of the {\it ab initio} calculations, and accounts somewhat for Breit and radiative corrections. 
We expect our calculations for the superheavy elements to be accurate to $\sim 0.1\%$.

\subsection{Nuclear dependence: volume isotope shift}

For the low $s$-levels of E119 and E120$^+$, we have found that there is a significant dependence 
on the root-mean-square nuclear charge radius $r_{rms}$. Our calculations were performed using 
a two-parameter Fermi distribution for the nuclear density. The values presented in the previous tables 
were performed with a half-density radius $c=8.0$~fm and 10-90\% width $t=2.0$~fm corresponding 
to a rms charge radius $r_{rms}\approx 6.42$~fm. Defining the volume isotope shift in terms of 
$r_{rms}$,
\begin{equation}
\frac{\delta E}{E}=k\frac{\delta r_{rms}}{r_{rms}} \, ,
\end{equation}
we have found the following values for $k$ for states $8s$ and $9s$ for E119 and E120$^+$ at the 
RHF level:
\begin{eqnarray}
{\rm E119}\  8s: \quad && k=-0.0243 \\
{\rm E119}\  9s: \quad && k=-0.0115 \\
{\rm E120}^+\ 8s: \quad && k=-0.0180 \\ 
{\rm E120}^+\ 9s: \quad && k=-0.00936 \, .
\end{eqnarray}

A table of values for $r_{rms}$ for nuclei $Z=119$ and $Z=120$ calculated in the nuclear Hartree-Fock-BCS 
approximation can be found in Ref. \cite{nuclear_rms}. The values range from around $r_{rms}=6.45$~fm 
to $r_{rms}=6.95$~fm for the very heavy isotopes. For $r_{rms}=6.90$~fm, we obtain at the RHF level the 
value $33495$\,cm$^{-1}$ for the removal energy for E119 8s. The difference between this value and 
that obtained with $r_{rms}=6.42$~fm, $\Delta =-59$\,cm$^{-1}$,  is comparable to the size of Breit 
and radiative corrections.

In principle, measurements of the spectra of different isotopes of superheavy elements may be used to 
get information about nuclear structure.

\section{Conclusion}

We have performed {\it ab initio} calculations of removal energies for the the low-lying $s$ and $p$ 
levels of the superheavy elements E119 and E120$^+$. 
Semi-empirical fitting was used to improve the accuracy of the calculations, accounting for neglected 
higher-order correlations as well as Breit and radiative corrections. The volume isotope shift was studied. 
The accuracy of our calculations is estimated to be on the order of $0.1\%$.

\acknowledgments

This research was supported by the Australian Research Council.

\end{document}